 \documentclass[final,3p,times]{elsarticle}
 \usepackage{graphics}
 \usepackage{graphicx}
 \usepackage{epsfig}

\usepackage{amssymb}
 \usepackage{amsthm}
\usepackage{verbatim}

\journal{Mathematical Biosciences}

\begin{document}

\begin{frontmatter}



\title{Spatial pattern formation induced by Gaussian white noise}

 \author[Politecnico]{Stefania Scarsoglio}
 \author[Politecnico]{Francesco Laio}
 \author[Virginia]{Paolo D'Odorico}
 \author[Politecnico]{Luca Ridolfi}
 \address[Politecnico]{Dipartimento di Idraulica, Trasporti ed Infrastrutture Civili, Politecnico di Torino, Torino, Italy}
\address[Virginia]{Department of Environmental Sciences, University of Virginia, Charlottesville, Virginia, USA}%

\begin{abstract}
The ability of Gaussian noise to induce ordered states in
dynamical systems is here presented in an overview of the main
stochastic  mechanisms able to generate spatial patterns. These
mechanisms involve: (i) a deterministic local dynamics term,
accounting for the local rate of variation of the field variable,
(ii) a noise component (additive or multiplicative) accounting for
the unavoidable environmental disturbances, and (iii) a linear
spatial coupling component, which provides spatial coherence and
takes into account diffusion mechanisms. We investigate these
dynamics using analytical tools, such as mean-field theory, linear
stability analysis and structure function analysis, and use
numerical simulations to confirm these analytical results.
\end{abstract}

\begin{keyword}
spatial patterns \sep noise-induced phenomena \sep Gaussian white noise


\end{keyword}

\end{frontmatter}

% \linenumbers
\section{Introduction}
\begin{figure} \centering
\includegraphics[width=12cm]{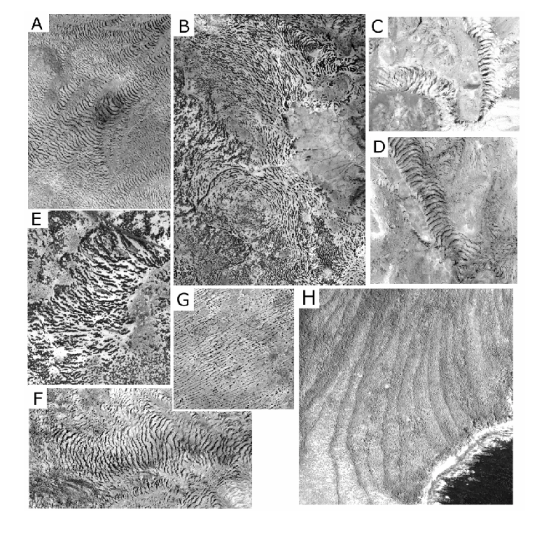} \caption{Example
of aerial photographs showing vegetation patterns (tiger bush).
(a) Somalia (9$^\circ$20'N, 48$^\circ$46'E), (b) Niger
(13$^\circ$21'N, 2$^\circ$5'E), (c) Somalia (9$^\circ$32'N,
49$^\circ$19'E), (d) Somalia (9$^\circ$43'N, 49$^\circ$17'E), (e)
Niger (13$^\circ$24'N, 1$^\circ$57'E), (f) Somalia (7$^\circ$41'N,
48$^\circ$0'E), (g) Senegal (15$^\circ$6'N, 15$^\circ$16'W), and
(h) Argentina (54$^\circ$51'S, 65$^\circ$17'W). Google Earth
imagery $\copyright$ Google Inc. Used with permission.}
\label{pattern_composite}
\end{figure}

Spatial patterns are widely present in different natural dynamical
systems. Their occurrence has been studied for quite a long time
with applications to different fields, including for example
hydrodynamic systems (e.g. Rayleigh-B\'{e}nard convection
\cite{Swift77,Wu95}) and biochemical and neural systems (see, for
instance, \cite{Shuai02,gluckman96}). In particular, a number of
environmental processes are known for their ability to develop
highly organized spatial features. For example, remarkable degrees
of coherence can be found in the spatial distribution of dryland
and riparian vegetation
\cite{Macfadyen50,Lefever97,Klausmeier99,Borgogno2009}), river
channels \cite{Rodriguez-Iturbe2001,Ikeda1989,Allen1985},
coastlines \cite{Ashton2001,Bird2000}, sand ripples and dunes
\cite{Lancaster1995}. Figure \ref{pattern_composite} shows an
example of natural spatial patterns around the world. These
patterns exhibit amazing regular configurations. Found over areas
of up to several square kilometers, they can occur on different
soils and with a broad variety of vegetation species and life
forms \cite{White1971,Montana1992,Eddy1999,Bergkamp1999}.

The study of patterns can offer useful information on the
underlying processes causing possible changes in the system. In
recent years, several authors have investigated the mechanisms of
pattern formation in nature, and their response to changes in
environmental conditions or disturbance regime. For example, in
the case of landscape ecology, these studies have related
vegetation patterns to the underlying eco-hydrological processes
\cite{Klausmeier99,Barbier07,Ridolfi07}, the nature of the
interactions among plant individuals \cite{Lefever97,Barbier07},
and the landscape's susceptibility to desertification under
different climate drivers and management conditions
\cite{vonHardenberg2001,dodorico05}.

Deterministic mechanisms of pattern formation have been widely
studied \cite{Turing1952,CrossHohenberg1993,Chandrasekhar1961}
with a number of applications to environmental processes
\cite{Murray2002,Borgogno2009,Lefever97,Manor2008,Couteron2001,Rietkerk2008,Kefi2007,Lefever2009}.
Stochastic models have only been developed more recently
\cite{Sagues2007,GarciaOjalvo1999}. They explain pattern formation
as a noise-induced effect in the sense that patterns can emerge as
a consequence of the randomness of the system's fluctuations.
These random drivers have often been related
\cite{Vandenbroeck1994a,VanDenBroeck97} to a symmetry-breaking
instability. They destabilize a homogeneous (and, thus, symmetric)
state of the system and determine a transition to an ordered
phase, which exhibits a degree of spatial organization. In the
thermodynamics literature these order-forming transitions are
usually referred to as non-equilibrium transitions, to stress the
fundamental difference in the role of noise with respect to the
classical case of equilibrium transitions, which exhibit an
increase in disorder as the amplitude of internal fluctuations
increases.

Here, we propose an overview of the main stochastic processes
related to the presence of Gaussian white noise, focusing on the
fundamental mechanisms able to induce spatial coherence. We
concentrate on Gaussian white noise because it provides a
reasonable assumption for the unavoidable randomness of real
systems -- the spatial and temporal scales of the Gaussian white
noise are much shorter than the characteristic scales over which
the spatio-temporal dynamics of the field variable are evolving --
and, therefore, it is typically adopted in stochastic modeling.
Moreover, the white noise assumption simplifies analytical and
numerical calculations. We call "patterned" a field that exhibits
an ordered state with organized spatial structures. This general
definition, including both periodic as well as multiscale
patterns, is often adopted in the environmental sciences, where
the number of different processes can prevent the organization of
the system with a specific wavelength. We define multiscale those patterns that are \emph{scale-free}
\cite{Manor2008b,vonHardenberg2010}, in the sense that spatial coherence emerges without showing a clear periodicity. Depending
on their behavior in time, patterns can be also classified as
steady or transient, on the basis of whether the spatial coherence
is constant in time or appears only temporarily with the tendency
to fade out with time. Steady patterns are here defined as statistically steady in time. This means that, once the steady
state is reached, the field variable can locally assume different
values, but the mean characteristics of ordered spatial structures remains the same. In the case of oscillating patterns the spatial
coherence instead fluctuates with time and patterns periodically
emerge and disappear.

In Section \ref{stochastic_modeling} a mathematical model of the spatio-temporal dynamics is introduced. We then consider two simple stochastic models (Sections \ref{additive_noise} and \ref{mul}), in order to clarify the interplay among three fundamental mechanisms: the local dynamics, the noise component and the spatial coupling. Pattern formation with temporal phase transition is described in Section \ref{pattern_phase_transition}. Concluding remarks are given in Section \ref{conclusions}. Analytical prognostic tools and numerical algorithms for pattern detection are treated in \ref{analytical_numerical}.

\section{Stochastic Modeling} \label{stochastic_modeling}
The spatio-temporal dynamics of the state variable, $\phi$, can be
expressed, at any point ${\bf r}=(x,y)$, as the sum of
four terms: (i) a function, $f(\phi)$, of local dynamics;
(ii) a multiplicative noise term, $g(\phi) \xi({\bf r},t)$; (iii)
a term, $D {\cal L}[\phi]$, accounting for the spatial
interactions with the other points of the domain, and (iv) an
additive random component $\xi_a({\bf r},t)$. Therefore, the
dynamics read

\begin{equation}
\frac{\partial \phi}{\partial t}=f(\phi)+g(\phi)  \xi({\bf r},
t)+ D{\cal L}[\phi] + \xi_a({\bf r},t),
\label{general_model}
\end{equation}
where ${\cal L}$ is an operator expressing the spatial coupling of
the dynamics, while $D$ is the strength of the spatial coupling.
The description of the spatio-temporal stochastic resonance and
coherence -- two mechanisms of noise-induced pattern formation
that need the cooperation of a temporal periodicity -- is not
included in this review. Thus, we will concentrate on the case of
dynamics in which the state of the system is determined by one
state variable, $\phi$, without time-dependent forcing terms.

The crucial point in the dynamical systems here investigated is
that pattern formation is noise-induced, i.e., it is due to random
fluctuations and does not occur in the deterministic counterpart
of the dynamics. In fact, these symmetry-breaking states vanish as
the noise intensity drops below a critical value depending on the
specific spatiotemporal stochastic model considered. For some
configurations these noise-induced transitions are re-entrant.
This means that the ordered phase is reached beyond a threshold
but is then destroyed if the noise intensity exceeds a higher
threshold. In these cases, the noise has a constructive effect
only when its intensity is within a certain interval of values.
Smaller or larger values are either too weak or too strong to
induce ordered structures. We consider a white (in time and space)
Gaussian noise with zero mean and correlation given by

\begin{equation}
\langle \xi({\bf r},t) \xi({\bf r'},t') \rangle = 2 s \delta({\bf r}-{\bf r'})\delta(t-t'),
\end{equation}
\noindent where $s$ is the noise intensity. We interpret the
Langevin equation (\ref{general_model}) in the Stratonovich sense.
In this case $\langle g(\phi) \xi(\phi) \rangle=s
\langle g(\phi)g'(\phi) \rangle$, where $g'(\phi)$ is the derivative of $g$ with respect to $\phi$. In contrast, under Ito interpretation, one has
$\langle g(\phi) \xi(\phi) \rangle=0$ \cite{Sagues2007,GarciaOjalvo1999}.

A number of mathematical models can be used to express the spatial
coupling in spatiotemporal dynamics.  We call
\emph{pattern-forming} those operators that, under suitable
conditions, are able to generate periodic patterns even without
noise. In contrast, non-pattern-forming operators are able to give
spatial coherence inducing multiscale patterns, without selecting
a clear dominant length scale. A typical example of
non-pattern-forming operator is the Laplacian,

\begin{equation}
{\cal L}[\phi]=\nabla^2 \phi=\frac{\partial^2\phi}{\partial
x^2}+\frac{\partial^2\phi}{\partial y^2}, \label{laplacian}
\end{equation}
which is widely used to represent the effect of the diffusion
mechanisms in a dynamical system. This operator accounts for
spatial interactions between a point of the domain and its nearest
neighbors, and is therefore considered as a short-range spatial
coupling.

A mathematical structure able to describe
pattern-forming couplings is instead the Swift-Hohenberg operator

\begin{equation}
\label{coupling-SH} {\cal SH}[\phi]=- (\nabla^2+k_0^2)^2\phi,
\end{equation}
where $k_0$ is a parameter corresponding to the wavenumber
selected by the spatial interactions. It should be noted that,
beside the effect of short-range interactions expressed by the
Laplacian operator $\nabla^2$, the biharmonic term, $\nabla^4$,
accounts for long-range interactions. Indeed, in a finite
difference discrete representation of equation (\ref{coupling-SH})
the biharmonic operator accounts for interactions with points of
the domain located next to the nearest neighbors. The
Swift-Hohenberg operator is one of the simplest types of coupling
able to account for both short and long range interactions and to
form periodic patterns. For this reason it has been widely adopted
in different applications \cite{CrossHohenberg1993,Sagues2007}.
The structure (\ref{coupling-SH}) was first introduced by
\cite{Swift77} to study the effect of hydrodynamic fluctuations in
systems exhibiting Rayleigh-B\'{e}nard convection
\cite{Chandrasekhar1961}. However, the interplay between short and
long range interactions expressed by (\ref{coupling-SH}) is a
recurrent  mechanism of pattern formation in nature. For example,
in landscape ecology, cooperative interactions for vegetation
growth -- such as mulching, shading, absence of biological crusts
\cite{Lejeune2004,Dodorico2007,Zeng2004,Borgogno2007,Joffre1993}
-- occur in the short range of plants' crown areas, while
inhibitory effects hindering vegetation establishment -- such as
competition for water and nutrients through the root system
\cite{Lefever2009,Lejeune2004,Barbier2006,Rietkerk2004} --
typically occur at larger distances.

\section{Additive noise} \label{additive_noise}
Consider the stochastic model

\begin{equation}
\label{proto_add+pattform} \frac{\partial \phi}{\partial t}=a \phi +D {\cal L}[\phi] + \xi_a({\bf r},t)
\end{equation}
where $\phi({\bf r},t)$ is the scalar field, $a$ is a parameter,
and $\xi_a({\bf r},t)$ is a zero-mean Gaussian white (in space and
time) noise with intensity $s_a$. Equation
(\ref{proto_add+pattform}) is the prototype model used to show how
patterns may occur in the absence of multiplicative noise (i.e.,
$g(\phi)=0$ in the general equation (\ref{general_model})) and of
a time-dependent forcing. We concentrate on linear deterministic dynamics to point out the fundamental mechanisms able to induce pattern formation, without invoking nonlinearities, which in this case do not substantially change the pattern properties. In so doing, we present very simple, though common and realistic, models of pattern formation. In this section we will
first study the case where ${\cal L}[\phi]$ is a pattern forming
spatial coupling.

\subsection{Pattern forming coupling} \label{additive+pattformcoupl}
The prototype
model is

\begin{equation}
\label{proto_add+pattform2} \frac{\partial \phi}{\partial t}=a \phi - D (\nabla^2+k_0^2)^2 \phi + \xi_a.
\end{equation}
The deterministic part of the dynamics does not generate patterns
for any value of $a$: if $a<0$ the system is damped to zero
without showing any spatial coherence, if $a$ is positive, no
steady states exist and the dynamics of $\phi$ diverge without
displaying any ordered spatial structures. Additive noise,
$\xi_a$, is able to keep the dynamics away from the homogenous
deterministic steady state even though in the underlying
deterministic dynamics $f(\phi)$ would tend to cause the
convergence to the homogenous state. In these conditions patterns
emerge and are continuously sustained by noise. These patterns are
noise-induced in that they disappear  and the homogeneous stable
state $\phi=0$ is restored if the noise intensity is set to zero.
\begin{figure}
\centering
\includegraphics[width=\columnwidth]{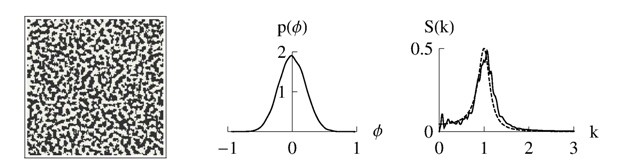}
\caption{Model (\ref{proto_add+pattform2}) at $t=100$, with
$a=-1$, $D=10$, $k_0=1$, and $s_a=0.5$. First panel: numerical
simulations of the field. Black and white tones are used for
positive and negative values of $\phi$, respectively. Second
panel: pdf of $\phi$. Third panel: azimuthal-averaged power
spectrum (solid: numerical simulations, dotted: structure
function).} \label{add_pf_sto}
\end{figure}
Figure \ref{add_pf_sto} reports some results from numerical
simulations, including the spatial field, the probability density
function (pdf) of $\phi$, and the azimuth-averaged power spectrum
and the structure function of the stochastic model
(\ref{proto_add+pattform2}). More details on the numerical methods
are provided in \ref{analytical_numerical}.  The results confirm
that a very clear and statistically stable pattern occurs
in spite of $a$ being negative: the noise component moves  the
dynamics away from the deterministic steady state $\phi_0=0$ and
allows the spatial differential terms to drive the field into a
 patterned state with wave length $2 \pi /k_0$.

This pattern-inducing role of the noise can be detected through
the structure function, as defined in \ref{struct_func}. Using
Equation (\ref{steady_struct}) one obtains

\begin{equation}
S_{st}({\bf k})=\frac{s_a}{D(-k^2+k_0^2)^2-a}.
\label{structure4}
\end{equation}

Thus, even for $a<0$ the steady state structure function has a
maximum at $k=k_0$ (see third row of figure \ref{add_pf_sto}, where the numerical power spectrum and the structure function are compared at steady state). This
result confirms that additive random fluctuations are able to
induce a stable pattern. When the noise is absent ($s_a=0$) the
steady state structure function is uniformly null and no patterns
form.

The normal mode stability analysis (see
\ref{normal_mode_analysis}) is unable to detect the occurrence of
patterns. In fact, according to this analysis no pattern should
emerge  when $a<0$. Similarly, the generalized mean field
technique (see \ref{gen_mean_field}) is unable to capture the
constructive role of additive noise.

The classical mean field technique (see \ref{clas_mean_field}) leads to

\begin{equation}
\label{SH_mf_m} \frac{{\rm d} \phi_i}{{\rm d}
t}=f(\phi_i)+g(\phi_i)\xi_i-D k_0^4 \phi_i -D
\left(\frac{20}{\Delta^4}- \frac{8k_0^2}{\Delta^2} \right)
(\phi_i- m)+ \xi_{a,i},
\end{equation}
\noindent where $\Delta$ is the spatial step (see
\ref{analytical_numerical}). This analysis shows that the order
parameter, $m$, does not change, i.e., the periodic patterns
induced by additive noise do not entail phase transitions. The
pdfs of the field variable, $\phi$, shown in Figure
\ref{add_pf_sto}, confirm the theoretical findings from the mean
field analysis. Indeed, one observes that the pdfs remain unimodal
and symmetrical at any time, with the mean at $\phi=0$ in spite of
the appearance of patterns.

\subsection{Non-pattern forming coupling}
In this section, we consider the same interplay between the local
deterministic component, $f(\phi)$, and the noise component
investigated in the previous section. However, we consider a
spatial coupling which is not able to select a specific
wavelength. The prototype model becomes

\begin{equation}
\label{proto_add+no+pattform} \frac{\partial \phi}{\partial t}=a \phi + D \nabla^2 \phi + \xi_a.
\end{equation}
The introduction of the additive random component allows one to
obtain very interesting patterned fields. The effect of additive
noise is even more surprising than  with a pattern-forming
coupling. In that case, (unsteady) patterns were in fact already
potentially present in the deterministic dynamics (see Section
\ref{additive+pattformcoupl}). In contrast, here the deterministic
dynamics does not reveal any transient spatial coherence.
\begin{figure}
\centering
\includegraphics[width=\columnwidth]{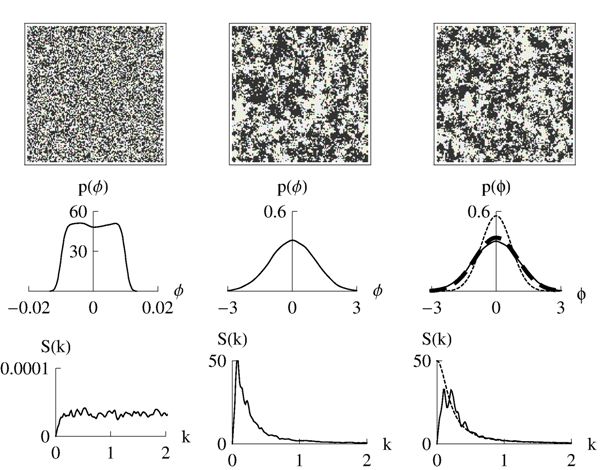}
\caption{Model (\ref{proto_add+no+pattform}) with $a=-0.1$, $D=2.5$, and $s_a=5$. The
columns refers to 0, 200, and 400 time units. First row: numerical simulations of the field. Second row: pdfs of $\phi$ (solid: numerical simulation, dotted: classic mean-field, dashed: corrected mean-field). Third row: azimuthal-averaged power spectrum (solid: numerical simulations, dotted: structure function).} \label{add_la_stoc}
\end{figure}

Considering the steady state structure function, one obtains

\begin{equation}
S({\bf k},t)=\frac{s_a}{Dk^2-a}
\end{equation}
\noindent which exhibits a maximum at $k=0$ (see third row of
figure \ref{add_la_stoc}, where the numerical power spectrum and
the structure function are shown at steady state). No specific
periodicity is selected, but a range of wave numbers close to zero
compete to give rise to multiscale patterns. However, since it is
difficult to predict the characteristics of these patterns only by
looking at the properties of the structure function, we have to
rely mostly on numerical simulations.

An example is reported in figure \ref{add_la_stoc}, where results
are shown in terms of the spatial field, the pdf, the
azimuth-averaged power spectrum and the structure function. As
expected from the analysis of the structure function, no clear
periodicity is visible, but many wave lengths are present. The
boundaries of the coherence regions are irregular and these
spatial structures fall then in the class of multiscale fringed
patterns, which are especially relevant in the environmental
sciences \cite{Manor2008b,vonHardenberg2010}. In fact, a number of environmental patterns exhibit a
spatial behavior very similar to the one shown in figure
\ref{add_la_stoc}. A typical example is the distribution of
vegetated sites in semi-arid environments \cite{Borgogno2009}.

In this case, other prognostic tools, such as the modal stability
analysis and the generalized mean field theory, fail to provide
useful indications. The pdf is unimodal (see the second row of the
figure \ref{add_la_stoc}) and its mean coincides with the basic
homogeneous stable state (i.e., $m=\phi_0=0$). Therefore, there is
no phase transition. This behavior follows the general rule that
Gaussian additive noise is unable to give rise to phase
transitions (i.e. changes of $m$) for any type of spatial
coupling.

Pattern formation induced by additive noise is usually introduced
in the scientific literature as a remarkable example of noisy
precursor near a deterministic pattern-forming bifurcation
\cite{Sagues2007}. The additive noise acts on a deterministic
system that exhibits a bifurcation point between a homogeneous
stable state and a stable patterned state (an example is the
Ginzburg-Landau model \cite{GarciaOjalvo1999,Sagues2007}). In this
case, the role of the additive noise is to unveil the intrinsic
spatial periodicity of the deterministic system even before
reaching the pattern-forming bifurcation.

\noindent This point of view suggests that a deterministic
bifurcation is necessary in order to have an additive noise
generating a pattern. The example with $f(\phi)=a \phi$ we have
just presented demonstrates, instead, that this is not necessarily
true. In this case there is no bifurcation since the dynamical
system diverges when $a>0$. Therefore, the existence of a
deterministic bifurcation is not  a necessary condition for
pattern formation. Patterns emerge as an effect of additive noise,
which unveils the capability of the deterministic component of the
dynamical system to induce transient periodic patterns also when
the asymptotic stable state is homogenous. Thus, noise exploits
this capability and hampers patterns to disappear.

\noindent Moreover, it should be noted that the presence of nonlinear components in the deterministic dynamics does not substantially change any of the previous results. The fine details of the patterns can change, but neither their stable occurrence nor their dominant wave length (if detectable) changes.

\section{Multiplicative noise}\label{mul}
The cooperation between multiplicative noise and spatial coupling
is based on two key actions: (i) the multiplicative random
component temporarily destabilizes the homogeneous stable state,
$\phi_0$, of the underlying deterministic dynamics, and (ii) the
spatial coupling acts during this instability, thereby generating
and stabilizing a pattern. The basic model is

\begin{equation}
\frac{\partial \phi}{\partial t}=f(\phi)+g(\phi)\xi({\bf r},t)+D {\cal L}[ \phi], \label{parrondo1}
\end{equation}
where, with respect to the general equation (\ref{general_model}), $\xi_a$ has been eliminated in order to isolate the role of
the multiplicative noise. $\xi$ is a zero-average Gaussian white
noise with intensity $s$. We indicate with $\phi_0$ the stable
homogeneous state of the system in the deterministic case. Namely,
$\phi({\bf r},t)=\phi_0$ is a homogeneous solution of
(\ref{parrondo1}) when $s=0$ (i.e., $f(\phi_0)=0$ because ${\cal
L}[ \phi]=0$ in homogeneous states). Moreover, we consider cases
where $g(\phi_0)=0$, so that the noise does not have the
possibility to destabilize the homogeneous steady state.

The analytical tools detecting the possible presence of the short
term instability are described in \ref{short_term_instab}. For
values of $s$ lower than a critical value, $s<s_c$, the state
variable $\phi({\bf x},t)$ experiences fluctuations about $\phi_0$
but noise does not play any constructive role. The system remains
blocked in the disordered phase and no patterns occur. Only
transiently, the spatial coupling might be able to induce patterns
that disappear as the system approaches its steady state.
Conversely, when the noise increases above a critical level,
$s>s_c$, the spatial term can take advantage of the noise-induced
short term instability and prevent that the displacement from the
homogeneous equilibrium state decays to zero. In this way, the
spatial coupling traps the system in a new ordered state,
maintaining the dynamics far from the state $\phi_0$. Equation
(\ref{parrondo1}) is here interpreted in the Stratonovich sense,
where $\langle g(\phi) \xi(\phi) \rangle=s \langle g(\phi)g'(\phi) \rangle$, while
under Ito's interpretation no short-term instability occurs, as
$\langle g(\phi) \xi(\phi) \rangle=0$ (see
\ref{short_term_instab}).

Other works have been proposed to describe the role of
multiplicative noise in pattern formation and phase transitions
\cite{Vandenbroeck1994a,VanDenBroeck97,Becker1994}, and to show how
multiplicative noise can induce periodic patterns
\cite{Parrondo96,GarciaOjalvo1999}. However, they all present
quite complicate non linear expressions to represent the local
dynamics and the noise terms, so that their physical
interpretation is not always straightforward. For example, when
$g(\phi_0)\neq0$, the noise plays a role similar to $\xi_a({\bf
r},t)$ in Eq. (\ref{proto_add+pattform}). Moreover, the results of
those models are qualitatively similar to those described in the
following sections, provided that the interplay
between short-term instability and spatial coupling remains the
same.

\subsection{Pattern forming coupling}\label{mult_pat}
\begin{figure}
\centering
\includegraphics[width=\columnwidth]{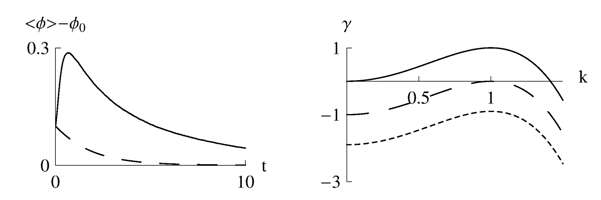}
\caption{(a) Behavior of $\langle \phi \rangle-\phi_0$ obtained as
ensemble average of $10^6$ realizations of the model
(\ref{proto_mult+pattform_temporal}) with $a=-1$. The initial
condition is $\phi=0.1$, $s=0.5$ and $s=5$ for the dashed and
solid curves, respectively. (b) Dispersion relation of the model
(\ref{proto_mult+pattform}), with $s=0.1, 1, 2$ (dotted, dashed and solid curves, respectively), $a=-1$,
$D=1$, and $k_0=1$.} \label{trans_proto_multi}
\end{figure}

To illustrate how pattern formation can be driven by multiplicative noise in the presence of a pattern-forming spacial coupling, we consider the model

\begin{equation}
\label{proto_mult+pattform} \frac{\partial \phi}{\partial t}= a
\phi - \phi^3 +\phi \xi - D(k_0^2+\nabla^2)^2 \phi,
\end{equation}
where $a$ is a negative number, the random component is modulated
by a function $g(\phi)=\phi$, and $\xi$ is a zero mean white
Gaussian noise, with intensity $s$. The local dynamics are
$f(\phi)=a \phi-\phi^3$,  where the nonlinear term $-\phi^3$ has
been introduced to avoid that the dynamics diverge. Indeed, the
linear term prevents the dynamics from relaxing to a homogeneous
steady state, thereby allowing the spatial terms to be different
from zero, while the nonlinear term prevents the divergence of the
dynamics (i.e., it ensures the convergence to a statistically
stable state). The deterministic homogeneous stable state,
obtained as a solution of $f(\phi_0)=0$, is $\phi({\bf
r},t)=\phi_0=0$. From the short-term instability (see
\ref{short_term_instab}) one has that $\phi_0$ is stable for
$s<-a$ and becomes instable for $s>-a$. To show this point, Figure
\ref{trans_proto_multi}a reports the time behavior of the ensemble
average of a number of numerically evaluated realizations of the
zero-dimensional stochastic model obtained eliminating the spatial
component from equation (\ref{proto_mult+pattform}), that is

\begin{equation}
\frac{{\rm d} \phi}{{\rm d} t}=f(\phi)+g(\phi)\xi(t)=a \phi -
\phi^3+\phi\xi(t). \label{proto_mult+pattform_temporal}
\end{equation}
It is evident from figure \ref{trans_proto_multi}a that the growth
phase appears only when $s>s_c=-a$ and at the beginning of
simulations (i.e., at short term), while the effect of the initial
perturbation disappears in the long run. Mathematically, the
short-term instability can be understood by observing that when
$\phi$ is close to zero the disturbance effect due to the
(multiplicative) noise tends to prevail on the restoring effect of
$f$. When $\phi$ grows, the leading term, $\phi^3$, prevails on
$g$ and the local dynamics, $f$, tend to restore the state
$\phi_0$.

Once the presence of a short-term instability has been detected,
the capability of the spatiotemporal stochastic model
(\ref{proto_mult+pattform}) to give rise to patterns can be
investigated through the stability analysis by normal
modes, see \ref{normal_mode_analysis}. The dispersion relation is

\begin{equation}
\gamma(k)=a+ s - D(k_0^2-k^2)^2,
\end{equation}
which provides the same threshold $s_c$ for the neutral stability
found with the short-term instability, while the maximum
amplification is for the wavenumber $k=k_0$ (see figure
\ref{trans_proto_multi}b). It follows that statistically steady
periodic patterns, with wave length $\lambda =2 \pi /k_{0}$,
emerge when the noise intensity exceeds the threshold $s_c=-a$.
Moreover, the critical value of the noise intensity for the
neutral stability, $s_c=[-a+D(k_0^2-k^2)^2]$, is confirmed by the
structure function (see \ref{struct_func}). The generalized
mean-field analysis of the most unstable mode predicts an
unconfined region of instability. This means that, if $s>s_c$, the
system is able to exhibit periodic patterns for any value of $D$.

\begin{figure}
\centering
\includegraphics[width=\columnwidth]{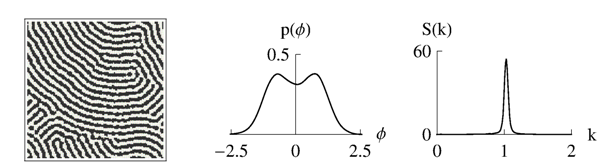}
\caption{Model (\ref{proto_mult+pattform}) at $t=100$, with $a=-1$,
$D=15$, $k_0=1$, $s=2.5$. First panel: numerical simulations of the field. Second panel: pdf of $\phi$. Third panel: azimuthal-averaged power spectrum.} \label{pattern_multi_proto_pf_super}
\end{figure}
The numerical simulation of the stochastic model
(\ref{proto_mult+pattform}) confirm these theoretical findings.
Figure \ref{pattern_multi_proto_pf_super} shows an example of
patterns emerging from the simulation. These patterns have the
same basic characteristics as those observed in the case of
additive noise (see figure \ref{add_pf_sto}). They are
statistically stable and exhibit a clear dominant wave length
corresponding to $k_0$. In this case the pdf of the field is weakly bimodal with the zero mean, demonstrating that no phase
transition occurs. However, there are some important differences
with respect to the case of additive noise. Firstly, the
boundaries appear to be more regular when patterns are induced by
multiplicative noise. Such aspect is also displayed by the power
spectrum of the field, which shows a more sharp peak at $k_0$ in
the case of multiplicative noise (compare the third rows of
figures \ref{add_pf_sto} and \ref{pattern_multi_proto_pf_super}).
This difference is due to the fact that multiplicative noise is
modulated by the local value of $\phi$, and this has the effect to
make the boundaries  of the pattern more regular because
the $\phi$ field is spatially correlated (from the definition
itself of patterned state). Another difference is that patterns
induced by additive noise exhibit more stable shapes than those
emerging as an effect of multiplicative noise. For example,
patterns shown in figure \ref{pattern_multi_proto_pf_super} seem
to evolve from a labyrinthine shape to a striped shape.
Overall, numerical simulations show that patterns induced by additive noise quickly reach their steady state without showing any transient temporal evolution, while those induced by multiplicative noise present a transient behavior, during which they modify their shape until the steady state is reached. The third difference is the possible occurrence of a weak bimodality
in the pdf of $\phi$ in the dynamics driven by multiplicative
noise (see figure \ref{pattern_multi_proto_pf_super}). The
presence of such bimodality depends on model structure, parameter
values, and field size; however, it generally remains weak.

Let's now look at the effect of nonlinear $g(\phi)$ terms on the
short-term instability. Consider equation
(\ref{proto_mult+pattform}) with $a<0$ and $g(\phi)=\phi^\alpha$.
If $\alpha>1$ no short-term instability occurs. This result is
explained interpreting the short-term behavior as a balance
between the tendency of $f(\phi)$ to restore the homogeneous
state, $\phi=\phi_0=0$, and the diverging action of $g(\phi)\cdot
\xi$. Since $\phi$ is close to zero, the power $\alpha>1$ of the
function $g(\phi)$ reduces the effect of the noise term, which
becomes unable to contrast the action of the leading term, $a
\phi$, of $f(\phi)$. Patterns occur only transiently and the field
then rapidly decays to the homogeneous state $\phi_0$.

\noindent Conversely, if $\alpha<1$, short-term instability occurs
in that the multiplicative component always overcomes the action
of $f(\phi)$ close to zero. The balance between $f(\phi)$ and
$g(\phi) \xi$ is reverted when $\phi$ moves away from $\phi_0=0$,
and this fact hampers the dynamical system to diverge. In this
case, patterns are statistically stable and exhibit the same dominant wave
length as those shown in figure
\ref{pattern_multi_proto_pf_super}.

\subsection{Non-pattern forming coupling}
We explore the capability of spatio-temporal models driven by multiplicative noise to generate patterns when a non-pattern forming spatial coupling is adopted. To focus on the role of the type of spatial operator, we consider the same model as in Section \ref{mult_pat}, but with a diffusive Laplacian operator, namely

\begin{equation}
\label{proto_mult+no_pattform} \frac{\partial \phi}{\partial t}=a \phi -\phi^3 +\phi \xi +D \nabla^2
\phi.
\end{equation}
where $\xi$ is a zero mean white Gaussian noise, with intensity
$s$, and equation (\ref{proto_mult+no_pattform}) is interpreted in
the Stratonovich sense. The model (\ref{proto_mult+no_pattform})
has a homogeneous deterministic stable state at $\phi_0=0$ and
exhibits short-term instability when $s>s_c=-a$ (see Section
\ref{mult_pat}). The  dispersion relation obtained with the linear
stability analysis reads

\begin{equation}
\label{disper_proto_no_pf} \gamma(k)=a + s - D k^2 .
\end{equation}
We can make the following three remarks. First, the value
$s=-a$ of the noise intensity marks the condition of marginal
stability. No unstable wave numbers occur when $s < -a$, while the
wave numbers lower than $\sqrt{(s+a)/D}$ become unstable if
$s>-a$. The threshold $s=-a$ coincides with the one obtained in
the short term analysis. Second, the strength, $D$, of the spatial
diffusive coupling impacts the range of unstable wavenumbers. In
particular, the unstable wave numbers decrease when $D$ increases,
consistently with the fact that the diffusive coupling introduces
spatial coherence in the random field. However, $D$ does not
impact the occurrence of instability, in that it depends only on
the noise intensity. Third, the most (linearly) unstable mode is
always $k_{max}=0$, for any $D$ and provided that $s>-a$. The
classic mean-field analysis predicts a phase transition for any
value of the spatial coupling, $D$. In other words, once the
critical threshold of noise intensity is exceeded, the system is
always able to move to a new ordered state.

\begin{figure} \centering
\includegraphics[width=\columnwidth]{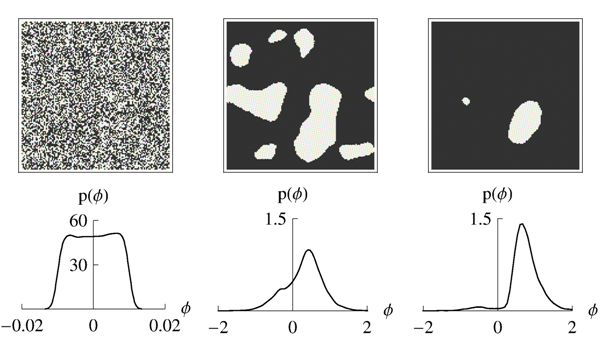}
\caption{Model
(\ref{proto_mult+no_pattform}) with $a=-1$, $s=2, D=5$. The columns refer to 0, 10, and 40
time units. First row: numerical simulations of the field. Second row: pdfs of $\phi$.} \label{pattern_multi_proto_no_pf}
\end{figure}

Figure \ref{pattern_multi_proto_no_pf} shows that the emergence of
a pattern with no clear periodicity. It evolves in time and tends
to disappear in the long term. The pdf of the field reveals that
patterns occur during a phase transition from the initial basic
state having order parameter $m=\phi_0=0$ to a new substantially
homogeneous state with $m \ne 0$. In particular, for the case
shown in figure \ref{pattern_multi_proto_no_pf} numerical
simulations give $m=0.8$ for $t>100$ time units.
Although with diffusive coupling both the cases of  additive and
multiplicative noise (not shown for sake of brevity) exhibit a
well defined peak at $k=0$, numerical simulations show a different
scenario (compare figures \ref{pattern_multi_proto_no_pf} and
\ref{add_la_stoc}). While the presence of additive noise produces
steady multiscale fringed patterns, multiplicative noise induces
only  smooth transient patterns. The reason of this different
temporal behavior is that, in the multiplicative case, the
diffusive operator is unable to maintain the system far from
homogeneous condition, in spite of the initial instability.
Indeed, to have steady patterns sustained by multiplicative noise
the presence of a pattern-forming spatial coupling is necessary.
When other types of spatial couplings are considered, they are
unable to block the system far from the homogeneous state. In
these cases, the spatial coupling interacts with the short-term
instability, as detected by the dispersion relation -- recall that
the stability analysis is performed on an equation that
approximates only the first stages of the ensemble average
dynamics -- but this interaction lasts only until the temporal
dynamics are able to sustain the instability. Thereafter, the
patterns undergo the same fate as the initial instability, i.e.,
they tend to disappear. In the long run the main legacy of the
spatial coupling is the phase transition (i.e., $m \ne 0$), though
for a homogeneous field. \noindent Similarly to the case discussed
in Section \ref{mult_pat}, the coherence regions are much smoother
than those observed with additive noise. Even in the this case,
this difference is due to the multiplicative nature of the noise,
which entails that the $g(\phi) \xi$ term is spatially correlated.

\section{Patterns with temporal phase transition} \label{pattern_phase_transition}
In this section we consider the extension to spatial systems of
noise-induced transitions in purely temporal systems \cite{Horsthemke1984}. Such transitions correspond to the occurrence
of steady state probability distributions whose modes are
different from the equilibrium states of the corresponding
deterministic system. A relevant case is represented by systems
exhibiting noise-induced bistability. In this case, suitable noise
intensities are able to generate pdfs with two modes even though
the deterministic dynamics have only one stable state. Two key
ingredients are needed to activate this type of stochastic
dynamics. First, a deterministic local kinetics, $f(\phi)$, which
tends to drive the dynamical system towards the steady state,
$\phi=\phi_0$. Second, a multiplicative random component that
tends to drive the state of the system away from $\phi=\phi_0$;
the intensity of this component is generally maximum at
$\phi=\phi_0$. As a result of the balance between deterministic
and stochastic components, bimodal probability distributions of
$\phi$ may emerge at steady state.

In spatiotemporal dynamical systems the spatial coupling could (i)
cooperate with the stochastic component to prevent the relaxation
imposed by the local dynamics and maintaining  the system away
from the uniform state, $\langle \phi \rangle=\phi_0$, and (ii)
give spatial coherence to the field creating a patterned state
where the coherent regions correspond to the two modes existing in
the underlying temporal dynamics.

The mechanism here discussed is sometimes called
\emph{entropy-driven pattern formation} \cite{Sagues2007} as the
dynamical system escapes from the minimum of the potential (i.e.,
$\phi=\phi_0$) because of the strength of  noise (which is an
entropy source). There are two main differences with respect to
the case presented in the previous sections: (i) patterns do not
result from a short-term instability, and (ii) they  emerge even
if the noise is interpreted according to Ito's rule.

\subsection{Model with $g(\phi_0)=0$}
It is interesting to consider what happens when no noise term is
present for  $\phi=\phi_0$, i.e. when $g(\phi_0)=0$. Indeed, in
this case the noise component is unable to unlock the system from
the deterministic stable state $\phi_0$ and to sustain the
pattern-forming effect of the spatial coupling by maintaining the
dynamics away from the homogeneous stable state, $\phi=\phi_0$.

\begin{figure} \centering
\includegraphics[width=\columnwidth]{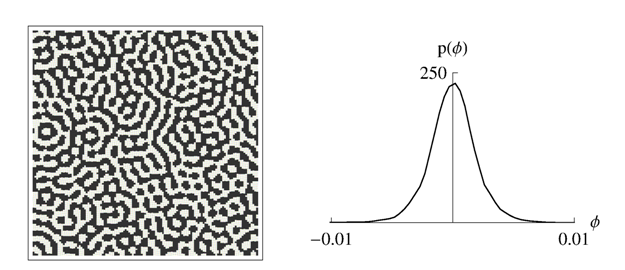}
\caption{Model (\ref{Ito}) under Ito
interpretation at $t=100$. The parameters are $a=0.001$, $D=10$, and
$s=1$.} \label{g_phi_0_0_Ito_SH}
\end{figure}

We consider the model

\begin{equation}
\label{Ito} \frac{\partial \phi}{\partial t}=-a \phi + \phi
(1-\phi) \xi (t) - D (k_0^2 + \nabla^2)^2 \phi,
\end{equation}
interpreted according to Ito. As in the case of section \ref{additive_noise} we concentrate on the case of linear local dynamics to show how patterns may emerge even without invoking nonlinearities in the underlying deterministic dynamics. The
purely temporal version of the model (\ref{Ito}) (i.e., ${\rm d}
\phi / {\rm d} t=-a \phi + \phi (1-\phi) \xi$) shows a
noise-induced transition for $s_c=4a$. The steady state pdf reads

\begin{equation}
\label{pdf_Ito_model}
p(\phi)=\frac{(1-\phi)^{\frac{a}{s}-2}}{\phi^{\frac{a}{s}+2}} {\rm
Exp} \left[ - \frac{a}{s(1-\phi)} \right] \hspace{1.5cm} \phi \in
]0,1[
\end{equation}
and has always a mode for $\phi \rightarrow 0$, while a second
mode occurs when $s > s_c$. The onset of bimodality in the model
(\ref{Ito}) is due to cooperation between the noise and the
natural boundaries at $\phi=0$ and $\phi=1$. When the noise is
sufficiently strong, the system tends to move away from $\phi=0$,
but the boundary at $\phi=1$ prevents the system from visiting the
whole real axis, and an accumulation of probability close to the
upper limit of the domain emerges.

\begin{figure} \centering
\includegraphics[width=\columnwidth]{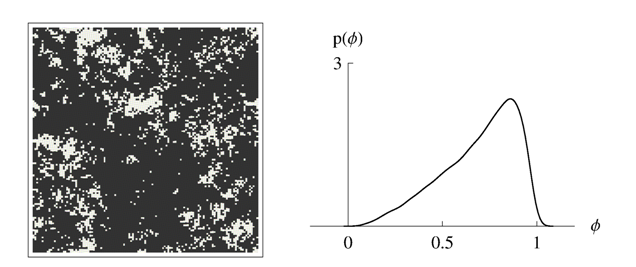}
\caption{Model (\ref{Ito_Laplacian}) at $t=300$ under Ito
interpretation. The initial conditions are given by uniformly distributed random numbers between [0.49, 0.51]. Black and white tones are used for the value intervals [0.5, 1] and [0, 0.5], respectively. The parameters are $a=0.001$, $D=25$, $s=40$.} \label{g_phi_0_0_Ito_Lapl}
\end{figure}
Since the noise is interpreted according to Ito, model (\ref{Ito})
does not present any noise-induced short term instability (i.e.,
$\langle g(\phi) \xi \rangle=0$). However, the spatial coupling is
able to exploit the temporal noise-induced transition and to show
patterns. In figure \ref{g_phi_0_0_Ito_SH} an example is
reported. In spite of the pdf of the temporal system displaying a
strong bimodality for high enough noise strength, the pdf of the
field is unimodal and centered in zero. The generalized mean-field
analysis is not able to provide further information.

\noindent Patterns can emerge also when in equation (\ref{Ito})
the diffusive spatial coupling is used in place of the
Swift-Hohenberg operator,

\begin{equation}
\label{Ito_Laplacian} \frac{\partial \phi}{\partial t}=-a \phi +
\phi (1-\phi) \xi (t) + D \nabla^2 \phi.
\end{equation}
Figure \ref{g_phi_0_0_Ito_Lapl} shows an example of these
patterns. For high enough values of the ratio $D/a$, the pdf is
unimodal and converges to $\phi=1$. However, the classic
mean-field analysis is not able to capture any phase transition of
the system.

\section{Conclusions} \label{conclusions}
We presented different stochastic mechanisms of spatial pattern
formation. They all describe spatial  coherence and organization
as noise-induced phenomena, in the sense that these patterns
emerge as an effect of the randomness of the system's drivers.

Additive noise plays a fundamental role when the deterministic
local dynamics tends to drive the field variable towards a uniform
steady state, while noise is able to maintain the dynamics  away
from the uniform steady state. The interaction of additive noise
with the spatial coupling provides a simple and realistic,
mechanism of pattern formation. In the presence of a
multiplicative component of adequate intensity the spatial
coupling exploits the initial instability of the system to
generate ordered structures, which in the absence of noise would
tend to disappear in the long run.

The stochastic models  presented here show how noise may play a
crucial role in pattern formation. However, most of the literature
on self-organized morphogenesis in the environment is based on
deterministic mechanisms. The limited application of stochastic
theories to environmental patterns is likely due to the fact that
most of the stochastic models use some specific (and complicated)
non linear terms both in the local deterministic dynamics and in
the multiplicative function, $g(\phi)$, of the noise component.
The use of these "ad hoc" functions limits the applicability of
these theories to process based environmental modeling. Thus, only
few studies have investigated the possible emergence of vegetation
patterns as a noise-induced effect. However,  because noisy
fluctuations -- such as those associated with fires, rain, or soil
heterogeneity  -- are a recurrent feature of environmental
drivers, their randomness can actually induce spatial coherence in
a number of environmental processes, as well as in problems related to front propagation under non-equilibrium conditions \cite{Armero1996,Panja2004,Sagues2007}.

%Therefore, stochastic models seem to offer some significant
%insights to the comprehension of pattern formation in
%ecohydrology.

 \appendix

\section{Analytical and numerical tools}\label{analytical_numerical}
The mathematical complexity of the spatio-temporal models of type
(\ref{general_model}) hampers general analytical solutions. For
this reason, several approximate analytical techniques have been
developed in order to obtain some indications of pattern
formation. We briefly describe the most important ones.

\subsection{Linear stability analysis by normal modes} \label{normal_mode_analysis}
When the occurrence of a dominant wavelength is the main symptom
of pattern formation, the first available prognostic tool is
provided by the normal mode linear stability analysis. This
analysis is based on the idea of disturbing the basic state of the
system with a hypothetical infinitesimal perturbation, and to
assess whether the perturbation grows in time (in which case
patterns have the possibility to emerge) or not. The analysis
involves three steps. Firstly, a deterministic equation for the
spatiotemporal dynamics of the ensemble average of the field
variable, $\langle \phi \rangle$, is determined and its
homogeneous steady state is found. In general, the multiplicative
random component can be expressed as $\langle g(\phi) \xi
\rangle=s \langle g_S(\phi) \rangle$, where $g_S(\phi)$ is a
function of the state variable. Applying Novikov's theorem
\cite{GarciaOjalvo1999}, for the case of Gaussian white noise
interpreted in the Stratonovich sense, we have $\langle g(\phi)
\xi \rangle=s\langle g(\phi)g'(\phi) \rangle$, i.e.
$g_S(\phi)=g(\phi)g'(\phi)$. Instead, with  Ito's interpretation
we have $\langle g(\phi) \xi \rangle=0$. Using equation
(\ref{general_model}), we find

\begin{equation}
\frac{\partial \langle\phi \rangle}{\partial t}=\langle
f(\phi)\rangle +s\langle g_S(\phi)\rangle+D {\cal L}[\langle \phi
\rangle]. \label{ensamble}
\end{equation}

The basic state, $\langle \phi \rangle=\phi_0$, is obtained
as the zero of equation (\ref{ensamble}) at steady state, i.e.
$f(\phi_0)+s g_S(\phi_0)=0$.

Equation (\ref{ensamble}) is then linearized, and Taylor's
expansion of equation (\ref{ensamble}) around $\phi=\phi_0$,
truncated to the first order, provides

\begin{equation}
\frac{\partial \langle\phi \rangle}{\partial t}=f'(\phi_0)\langle
\phi \rangle + s g_S'(\phi_0)\langle \phi \rangle+D {\cal
L}[\langle \phi \rangle], \label{lin_ensamble}
\end{equation}
where $f'(\phi_0)=\left.\frac{{\rm d}f(\phi)}{{\rm
d}\phi}\right|_{\phi=\phi_0}$ and $g_S'(\phi_0)=\left.\frac{{\rm
d}g_S(\phi)}{{\rm d}\phi}\right|_{\phi=\phi_0}$. The basic state
$\langle \phi \rangle=\phi_0$ is perturbed (third step) by adding
an infinitesimal harmonic perturbation

\begin{equation}
\langle \phi \rangle=\phi_0+\hat{\phi} e^{\gamma t+i {\bf k} \cdot
{\bf r}}, \label{linstab}
\end{equation}
with $\hat{\phi}$ being the perturbation amplitude, $\gamma$ the
growth factor, $i=\sqrt{-1}$ the imaginary unit, ${\bf k}=(k_x,
k_y)$ the wave number vector of the perturbation, and ${\bf
r}=(x,y)$ the coordinate vector. If equation (\ref{linstab}) is
inserted in (\ref{lin_ensamble}), one obtains the so-called
\emph{dispersion relation}

\begin{equation}
\gamma(k)=f'(\phi_0)+s g_S'(\phi_0)+D h_{\cal L}(k),  \label{disp}
\end{equation}
where $h_{\cal L}(k)$ is a function of the wave number $k=|{\bf
k}|$ which depends on the specific form of spatial coupling $\cal
L$ considered. From equation (\ref{disp})  the threshold value  of
the noise intensity is easily obtained by setting the marginal
condition, $\gamma=0$

\begin{equation}
s_c=-\frac{f'(\phi_0)+D h_{\cal L}(k)}{g_S'(\phi_0)},
\label{marg_stab}
\end{equation}
c When $s>s_c$, the growth factor, $\gamma$, is positive and
spatial patterns may occur.

\subsection{Short-term instability} \label{short_term_instab}
A second tool can be used to assess the possible short term
instability in the dynamical system. The transient instability is
important because it tends to move the dynamics away from the
basic state $\langle\phi\rangle=\phi_0$. If this phenomenon is
accompanied by a suitable spatial coupling, the system can be
trapped in a new ordered state. The first steps of the stability
analysis are the same as those described in
\ref{normal_mode_analysis}, and lead to Eq. (\ref{ensamble}).
Since we are here interested in the initial evolution of small
displacements from $\phi_0$, we can assume $\langle f(\phi)
\rangle\approx f(\langle \phi \rangle)$ and $\langle g_S(\phi)
\rangle\approx g_S(\langle \phi \rangle)$. For the same reason, we
can also neglect the spatial gradients of the fluctuations
assuming that  in the short term they are small. Equation
(\ref{ensamble}) can be approximated at any point of the field as
\cite{Sagues2007}

\begin{equation}
\frac{d \langle \phi \rangle}{dt}\approx f(\langle\phi\rangle)  +
s  g_S(\langle \phi \rangle) =f_{eff}(\langle\phi\rangle), \label{macroscopic1}
\end{equation}
where $f_{eff}$ is often indicated as the
effective kinetics. The short term stability
analysis of the state $\phi_0$ by equation (\ref{macroscopic1})
concerns only the temporal dynamics at a generic point of the
field.

Equation (\ref{macroscopic1}) clearly shows that noise can
destabilize the deterministically stable state $\phi_0$. This can
happen in two possible ways: (i) when the roots of
$f_{eff}(\langle \phi \rangle)=0$ do not coincide with those of
$f(\langle \phi \rangle)=0$ or (ii) when the state $\phi_0$
remains a zero of the r.h.s. of equation (\ref{macroscopic1}), but
sufficiently high noise intensities destabilize this state an
unstable one; this occurs when the following condition is met

\begin{equation}
\label{cond_stab} \left.\frac{{\rm d} f_{eff}}{{\rm d}
\langle\phi\rangle} \right|_{\phi_0} \geq 0.
\end{equation}
In both cases noise has to be multiplicative for a transition to
occur. The noise threshold $s_c$ can be obtained by setting the
inequality in (\ref{cond_stab}) equal to zero. The condition
(\ref{cond_stab}) derives from the first-order truncated Taylor
expansion of the function $f_{eff}$ around $\phi_0$, which yields

\begin{equation}
\frac{d \langle \phi \rangle}{dt}\approx \left.\frac{{\rm d} f_{eff}}{{\rm d} \langle\phi\rangle}
\right|_{\phi_0} \langle \phi \rangle.
\end{equation}
When $f_{eff}(\phi_0)=0$, the sign of the coefficient of $\langle
\phi \rangle$ on the r.h.s. of the previous relation determines
the stability of small perturbations around $\phi_0$.

When the noise is additive the effective kinetics are
$f_{eff}(\langle \phi \rangle)=f(\langle \phi \rangle)$. Thus,
the stable states of (\ref{macroscopic1}) are the same as those of
the deterministic counterpart of the process.

It is also worth to stressing the impact of the type of noise and
of its interpretation. While in the Stratonovich case,
$g_S(\phi)=s g(\phi)g'(\phi)$, using Ito interpretation we have
$g_S(\phi)=0$. Therefore, the noise-induced instability is
possible only in the Stratonovich interpretation of the Langevin
equation (\ref{general_model}) and it cannot occur when Ito's
framework is adopted.

\subsection{Structure function}\label{struct_func}
The  presence of patterns modifies the correlation structure of
the field. Instead of considering the correlation function, this
method analyzes its Fourier transform in space, which is known
with the name of structure function and defined as $S({\bf k},
t)=\langle \hat{\phi}({\bf k},t) \hat{\phi}({\bf -k},t) \rangle$,
where $\hat{\phi}({\bf -k},t)$ is the Fourier transform of
$\phi({\bf r},t)$ and $\textbf{k}=(k_x, k_y)$ is the wave number
vector. The structure function is therefore equal to the power
spectrum of the field $\phi$. The first-order temporal derivative
of the structure function reads

\begin{eqnarray}
\label{structure1}
\frac{\partial S({\bf k}, t)}{\partial t}= \left\langle \frac{\partial \hat{ \phi}({\bf
k},t)}{\partial t} \hat{\phi}({\bf -k},t) \right\rangle+
\left\langle \frac{\partial \hat{ \phi}({\bf -k},t)}{\partial t}
\hat{\phi}({\bf k},t) \right\rangle.
\end{eqnarray}
By taking the Fourier transform of the linearized version of Eq.
(\ref{general_model}) (with $F(t)=0$), which is obtained by
following similar steps to those leading to equation
(\ref{lin_ensamble}), and considering the noise terms as white in
space, we obtain

\begin{equation}
\frac{\partial \hat{\phi}({\bf k},t)}{\partial
t}=f'(\phi_0)\hat{\phi}({\bf k},t)+ g'(\phi_0)\hat{\phi}({\bf
k},t)\xi(t)+\xi_a(t)+ D h_{\cal L}(k)\hat{\phi}({\bf k},t),
 \label{ftransform2}
\end{equation}
where $h_{\cal L}(k)$ is the same operator already defined in
\ref{normal_mode_analysis}. By substituting Eq.
(\ref{ftransform2}) into Eq. (\ref{structure1}) and using
Novikov's theorem \cite{GarciaOjalvo1999} to express the terms
$\left<\hat{\phi}({\bf k},t)\hat{\phi}(-{\bf k},t)\xi (t) \right>=
s S({\bf k}, t)$ and $\left<\hat{\phi}(\pm {\bf k},t)\xi_a (t)
\right>=s_a$ (where $s_a$ is the intensity of the additive white
Gaussian noise), we obtain

\begin{equation}
\frac{\partial S({\bf k},t)}{\partial t}=2\left[f'(\phi_0)+D
h_{\cal L}(k) +g'(\phi_0)s \right]S({\bf k},t)
+2 s_a \label{structure4}
\end{equation}

At steady state the structure function reads

\begin{equation}
S_{st}({\bf k})=-\frac{s_a}{\left[f'(\phi_0)+D h_{\cal L}(k)
+g'(\phi_0) s \right]}. \label{steady_struct}
\end{equation}
Eq. (\ref{steady_struct}) can be investigated to understand if
periodic patterns, corresponding to a maximum of the structure
function for wave numbers $k$ different from zero, are expected to
appear. Equation (\ref{steady_struct}) shows that the additive
noise is fundamental to have a non-null steady-state structure
function.

\subsection{Mean-field analysis}
The mean-field theory is typically used to provide an approximated
solution of stochastic partial differential equations for
spatially-extended systems. The method is valuable mainly for a
qualitative analysis of (stochastic) spatiotemporal dynamics
\cite{Vandenbroeck1994a,VanDenBroeck97,Buceta03b,Porporato04}.

The mean field technique adopts a finite difference representation
of the stochastic spatiotemporal dynamics (\ref{general_model})

\begin{equation}
\label{syst_mf} \frac{{\rm d} \phi_i}{{\rm d} t}=f(\phi_i)+g(\phi_i) \cdot
\xi_i(t)+D \cdot l(\phi_i, \phi_j)+h(\phi_i)\cdot
F(t)+\xi_{a,i}(t),
\end{equation}
\noindent where $\phi_i$, $\xi_i$, and $\xi_{a,i}$ are the values
of $\phi$, $\xi$, and $\xi_a$ at site $i$, respectively, $i$ runs
across all the cells of the discretized domain, and $j \in nn(i) $
refers to the neighbors of the $i$-th site involved in the
discretized representation, $l(\phi_i,\phi_j)$, of the specific
spatial coupling considered. A general expression for
$l(\phi_i,\phi_j)$ is

\begin{equation}
l(\phi_i,\phi_j)=w_i \phi_i+ \sum_{j\in nn(i)} w_j \phi_j,
\end{equation}
where the number of neighbors, $nn(i)$, and the weighting factors
$w_i$ and $w_j$ depend on the specific finite difference scheme
adopted to numerically approximate the spatial operator
\cite{Strikwerda04}. The analytical solution of equation
(\ref{syst_mf}) is hampered by the fact that the dynamics of
$\phi_i$ are coupled to that of the neighboring points. To
circumvent this issue, the mean field approach assumes that (i)
the variables $\phi_j$ can be approximated by their local ensemble
mean, $\langle \phi_j \rangle$, and (ii) there is a relation
linking $\langle \phi_j \rangle$ to the ensemble average, $\langle
\phi \rangle$.

\subsubsection{Generalized mean-field theory} \label{gen_mean_field}
To study the  stability of the homogeneous steady state with
respect to periodic patterns, the pattern is approximated by a
harmonic function,

\begin{equation}
\label{mean_cos} \langle \phi_j \rangle=\langle \phi \rangle
\cos[{\bf k} \cdot ({\bf r}_i-{\bf r}_j)],
\end{equation}

\noindent where ${\bf k}=(k_x, k_y)$ is the wave number vector.
The function $l(\phi_i,\phi_j)$ in equation (\ref{syst_mf}) is
approximated as

\begin{equation}
\label{mean_field_k} l(\phi_i,\phi_j) \approx l_h(\phi_i, \langle
\phi \rangle, k_x, k_y)
\end{equation}
where $l_h(\cdot)$ depends on the
spatial coupling considered.

Under the assumption (\ref{mean_field_k}), the dynamics
(\ref{syst_mf}) of $\phi_i$ become independent of those of the
neighboring points. Thus, it is possible to determine exact
expressions for the steady-state probability distributions
$p_{st}(\phi; \langle \phi \rangle, k_x, k_y)$ of $\phi$. The
self-consistency condition

\begin{equation}\label{self_consistency}
\langle \phi \rangle=\int_{-\infty}^{+\infty} \phi \
p_{st}(\phi;\langle \phi \rangle, k_x, k_y) \ {\rm
d}\phi=F\bigl( \langle \phi \rangle, k_x, k_y \bigr)
\end{equation}
can be used to obtain the unknown $\langle \phi \rangle$ as a
function of $k_x$ and $k_y$.

The occurrence of solutions of equation (\ref{self_consistency})
different from $\langle \phi \rangle=\phi_0$ (where $\phi_0$ is
the uniform steady state of the system) corresponds to the loss of
stability of the uniform steady state with respect to periodic
perturbations. It is expected that this loss of stability takes
place only for some specific value of the wave numbers $k_x$ and
$k_y$.

Other noise-induced phenomena can be investigated with this
method. Indeed, spatial pattern formation is not the only
interesting noise-induced effect. In fact,  modifications of other
statistical descriptors of the field can be relevant, too.
Modifications of order parameters are known as \emph{phase
transitions}. In particular, when the spatio-temporal average,
$m$, of the state variable at steady state is different from the
 homogeneous steady state value, a phase transition occurs. Non-equilibrium phase transitions are induced by the random forcing.
 The occurrence of non-equilibrium phase transition is neither a necessary nor a
sufficient condition for noise-induced pattern formation.
Non-equilibrium phase transitions imply that noise is able to
change the value of the order parameter, but not that ordered
geometrical structures necessarily emerge. Conversely, we have
shown that noise-induced patterns may emerge even when $m$ remains
unchanged with respect to the disordered case (i.e., no phase
transition occur).

\subsubsection{Classic mean-field theory} \label{clas_mean_field}
The classic mean field theory can be presented as a simplified
version of the generalized mean field. More specifically, it is
assumed that all cells have the same mean, which coincides with
the spatiotemporal mean of the field. Namely, ${\bf k}=0$ in
equation (\ref{mean_cos}), i.e. $\langle \phi_j \rangle=\langle
\phi \rangle=m$. In this case, equation (\ref{self_consistency})
becomes

\begin{equation}\label{self_consistency_m}
\langle \phi \rangle=\int_{-\infty}^{+\infty} \phi \ p_{st}(\phi; \langle \phi \rangle) \ {\rm
d}\phi=F\bigl(\langle \phi \rangle \bigr).
\end{equation}
The change in the number of solutions of equation (\ref{self_consistency_m}) indicates the existence of a phase transition. The focus of this
analysis is not on the appearance of periodic patterns but only on
the occurrence of phase transitions. The effectiveness of this
standard mean field approximation can be improved by expressing
the values of $\phi_j$ in the neighborhood of point $i$ as the
average between the spatiotemporal mean and the local value of
$\phi$ at point $i$, namely $\phi_j \approx 1/2(\langle \phi
\rangle +\phi_i)$. This correction of the mean field approximation
accounts for the dependence of $\phi_j$ on the local conditions
\cite{Sagues2007}.

The analytical tools here described provide in general some
insights into pattern formation, but the definitive way to study
the noise-induced pattern formation is to numerically simulate the
dynamics and assess the emergence of spatial coherence  through a
comparison with homogeneous or disordered states of the system. In
fact, the linear stability analysis by normal modes and the
short-term instability analysis tend to fail when the noise is
 additive (i.e. $g=0$ in Eq. (\ref{general_model})). In this
case,  both analyses predict stable configurations for any noise
intensity and any strength of the spatial coupling. On the other
hand, when only multiplicative noise is present (i.e. $\xi_a=0$ in
Eq. (\ref{general_model})), the structure function does not
provide any information on pattern formation, as $\phi$ tends to
remain equal to zero.

The typical numerical approach is to discretize the continuous
spatial domain using a regular Cartesian lattice with spacing
$\Delta x=\Delta y=\Delta$. Here we consider a two-dimensional
square lattice with 128x128 sites and $\Delta=1$. The original
stochastic partial differential equation (\ref{general_model}) is
then transformed into a system of coupled stochastic ordinary
differential equations as in Eq. (\ref{syst_mf}). \noindent
Infinitely vast random fields are generally approximated
numerically in a satisfactory way by periodic boundary conditions,
which have been used also in this study. Moreover, unless it was
otherwise specified, the initial conditions used in the
simulations were uniformly distributed random numbers between
[-0.01, 0.01]. Numerical simulations were carried out with the
Heun's predictor-corrector scheme
\cite{VanDenBroeck97},\cite{Sagues2007}. The pdf is obtained from the spatial distribution of $\phi$ at fixed time and is numerically evaluated at $100$ equally spaced intervals, $\Delta \phi$, that cover the range of $\phi$ values.


%% References


%\bibliographystyle{elsarticle-num}



\end{document}